\documentclass[pra,reprint, nofootinbib,superscriptaddress]{revtex4-1}
\usepackage{graphicx}
\usepackage[colorlinks=true,linkcolor=blue,citecolor=red,urlcolor=blue]{hyperref}
\usepackage{amsmath, amssymb}
\usepackage{bm}
\usepackage{subfigure}
\usepackage{color}
\usepackage{physics}
\usepackage{soul,xcolor}

\newcommand{\be}{\begin{equation}}
\newcommand{\ee}{\end{equation}}

\begin{document}
\title{Angular Momentum Supercontinuum from Fibre Rings} 
\author{Calum Maitland}
\email{cm350@hw.ac.uk}
\affiliation{Institute of Photonics and Quantum Sciences, Heriot-Watt University, Edinburgh EH14 4AS, UK} 
\affiliation{School of Physics and Astronomy, University of Glasgow, Glasgow G12 8QQ, United Kingdom}
\author{Fabio Biancalana}
\affiliation{Institute of Photonics and Quantum Sciences, Heriot-Watt University, Edinburgh EH14 4AS, UK}

\begin{abstract}

Broadband, coherent light carrying optical angular momentum is of potential utility for a variety of classical and quantum communication applications, but at present few such sources exist. We study the generation of supercontinua in a ring array of coupled optical fibres. Short pulses carrying discrete angular momentum undergo soliton fission, spontaneously breaking azimuthal symmetry. This results in a train of pulses with a broadband frequency spectrum as well as a non-trivial angular momentum distribution. These spatio-temporal solitary waves, localised around a single fibre core, emit an unusual form of resonant radiation which can be present even in the absence of intrinsic higher order dispersion, being induced by the lattice dispersion of the ring array. We explore how the coupling properties between fibre cores affect the resulting supercontinuum, in particular how mildly twisting the array can effectively manipulate its angular momentum content and resonant frequencies through the induced Peierls phase.

\end{abstract}

\date{\today}
\maketitle

\section{Introduction}

Optical fibres have been used as platforms for supercontinuum generation for decades, primarily because their tight optical confinement provides the strong nonlinearity required for huge spectral broadening \cite{Lin1976,  Morioka1993, Mori2003, Dudley2010}. Multicore fibre structures are increasingly being studied as supercontinuum sources as they offer further mechanisms for dispersion management to extend the spectral range of the output \cite{XFang2012, NWang2018}. Meanwhile, the nonlinear optical properties of light carrying angular momentum (AM) are just beginning to be explored \cite{Gordon2009, Lanning2017, Pereira2017, Offer2018, Grigoriev2018}, in particular the possibility of generating broadband supercontinua as recently demonstrated by Prabhakar \textit{et. al.} \cite{Prabhakar2019}. Twisted ring arrays of fibres and other helical waveguides have been developed to explore new ways to manipulate optical AM in a discrete fashion \cite{Roth2018, Zannotti2017}. Optical AM may be exploited for a plethora of applications including astronomical observation \cite{Foo2005, Swartzlander2008, Elias2008}, classical \cite{Bozinovic2013, Zhu2018} and quantum \cite{Groblacher2006, Bouchard2018} communication, optomechanics \cite{Padgett2011, Shi2016} and microscopy \cite{Furhapter2005}. However, not many broadband and coherent optical sources   of AM exist and their development is becoming an active field of study, with several metasurface designs being realised in recent years \cite{Liu2016, Xia2019, Zhou2019, Tian2019}. In this work we examine how such structures may generate supercontinua when pumped with intense pulses of light, and how the coupling properties of the array affect the supercontinuum's frequency and AM content. 

We consider light propagating through a ring array of step index fibres, embedded in homogeneous  cladding material. Light can be exchanged between adjacent fibre cores though evanescent coupling due to the overlap of their fundamental modes. Hence we use a coupled mode approach, in which we consider only the complex amplitude of the fundamental guided mode within each core. The time-independent nonlinear properties of such multicore arrays have been widely studied before \cite{Turitsyn2012, Hadzievski2015, Martinez2015, Rubenchik2015, Chekhovskoy2016}, with and without an extra central core which is not considered here. In addition the fibre cores may be twisted around the cladding rod's central axis at a uniform rate, which can greatly change the supercontinuum spectra even for very moderate twist rates.  In previous work we have explored modulation instability in this system and how this enables light with different AM to be generated from continuous wave pumps with no AM \cite{Maitland2019}. Here we study how short, intense pulses undergo fission and induce supercontinua across the discrete angular momenta available. The generation of different AM from pump pulse with single-valued AM is the result of {\em angular symmetry breaking}, arising from instabilities driven by the nonlinearity.

Throughout this article we work in dimensionless units following the prescription in \citep{Agrawal2013}, with our propagation coordinate $z$ related to physical distance $\tilde{z}$ through the dispersion length, $z = \tilde{z}/L_D = \tilde{z} |\beta_2| / {T_0}^2$ given an input pulse duration $T_0$ and second order dispersion coefficient $\beta_2 \equiv [{\partial_\omega}^2 \beta]_{\omega_{0}}$, where $\beta(\omega)$ is the linear propagation constant of waves at frequency $\omega$ and $\omega_{0}$ is the central input pulse frequency. We work in a frame which rotates with the fibre twist such that the cores are fixed in position, and which is comoving with the incident pulse (which is travelling at the group velocity $v_g = 1 /[\partial_\omega \beta]_{\omega_{0}}$) by using a dimensionless time coordinate $t = (\tilde{t} - \frac{\tilde{z}}{v_g})/T_0$, where $\tilde{t}$ is the physical (laboratory) time.  We express the complex amplitude of the fundamental mode of the $n^{th}$ fibre core in dimensionless form, $A_n = \sqrt{\gamma L_D} E_n$, where $E_n$ is the physical electric field amplitude and $\gamma$ is the Kerr nonlinear coefficient. Light propagation through the fibre array is then described in the coupled mode limit by a series of nonlinear Schr\"{o}dinger equations in standard form (assuming anomalous dispersion $\beta_2 < 0$):
\begin{equation} \label{eq: coupledNLSE}
\begin{split}
i \partial_z A_n = &-\frac{1}{2}{\partial_t}^2 A_n+ \frac{i\beta_3}{6}{\partial_t}^3 A_n - {|A_n|}^2 A_n\\
&- \Delta \left(\exp(-i \phi) A_{n+1} + \exp(i \phi) A_{n-1} \right).
\end{split}
\end{equation}
%
Here $\beta_3 \equiv  L_D [\partial_\omega \beta]^3_{\omega_{0}}  / {T_0}^3$  is a dimensionless third order dispersion parameter, $\Delta$ is the nearest neighbour coupling rate between adjacent fibre cores and $\phi$ is an effective Peierls phase induced by the twisting of the array around its central axis \cite{Longhi2007b} (inversely proportional to the twist period). We define the coupling rate through the integral \cite{Longhi2007b}
\begin{equation} \label{eq: delta}
\Delta \equiv \frac{2\pi L_D}{\lambda_0} \int d \mathbf{r} u^*(\mathbf{r}-\mathbf{r}_{n+1}) n_c(\mathbf{r}-\mathbf{r}_n) u(\mathbf{r}-\mathbf{r}_n) 
\end{equation}
given the refractive index profile $n_c$ of a single step-index fibre core surrounded by cladding, fundamental mode profile $u(\mathbf{r})$, $\lambda_{0}\equiv 2\pi c/\omega_{0}$ the initial pulse's central wavelength and $\mathbf{r}_n$ the spatial coordinates of the centre of the $n^{th}$ fibre core.

We introduce the discrete angular coordinate $\theta_n \equiv 2 \pi n /N$ which labels the azimuthal position of fibre core $n = 1, 2, ... N$, where $N$ is the total number of cores in the ring array. Assuming only the fundamental modes are excited, the phase of the electric field on each core is well defined, allowing a quasi-AM to be defined which modifies this phase by a factor $\exp(i l \theta_n)$ where $l \in [-N/2+1, N/2]$ for $N$ even or $l \in [-(N-1)/2, (N-1)/2]$ for $N$ odd is the AM integer order. This is related to but distinct from optical orbital angular momentum (OAM), which takes unrestricted integer values and is defined for spatially continuous electric fields. With no higher-order dispersion $\beta_3=0$, eq. \eqref{eq: coupledNLSE} possesses discrete ``vortex'' soliton solutions with winding number $m$ of the form 
\begin{equation} \label{eq: DVsoliton}
\overline{A}_n(z,t) = B \sech{\left(B t \right)} \exp(i \overline{\beta} _m z  + i m \theta_n)
\end{equation} 
where $\overline{\beta} _m = B^2/2 + 2 \Delta \cos( 2 \pi m/N - \phi) $ is the soliton's propagation constant. 

\section{Soliton Fission \& Angular Symmetry Breaking}

\begin{figure}[h] 
\centering
\includegraphics[width=\linewidth]{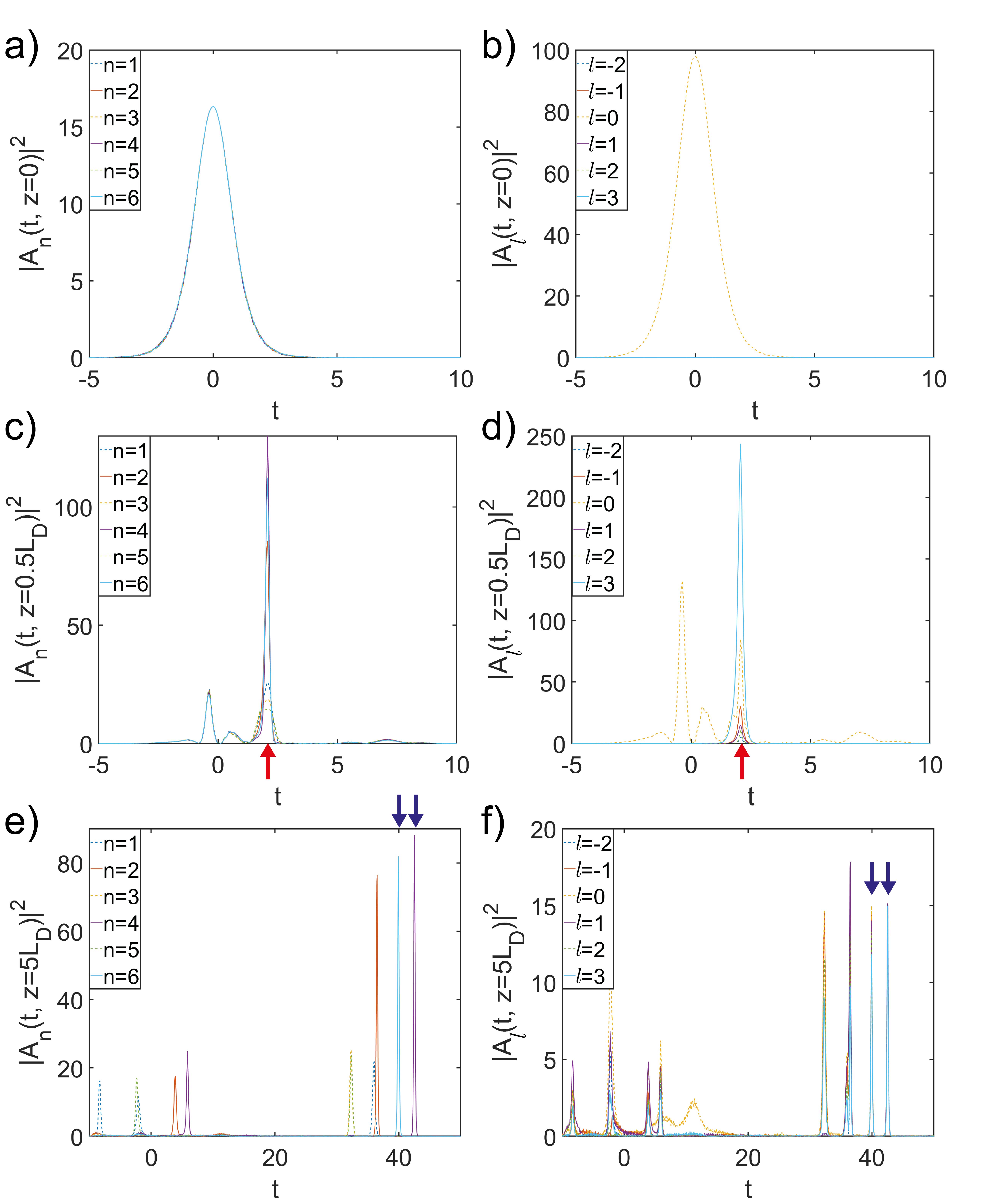}
\caption{\small{The azimuthal symmetry breaking process is illustrated by the intensity distribution over time, as viewed in the fibre index (left) and AM (right) basis. The intense ($A_n(0, t) = 4 \overline{A}_n(0,t)$, $B=1$) pulse at the input (panels \textbf{a)} and \textbf{b)}) with AM $m=0$ collapses and undergoes fission. The brightest emerging pulse experiences instability and symmetry breaking at the most intense point, indicated by red arrows,  after $z=0.5L_D$ populating the other AM (panels \textbf{c)} and \textbf{d)}). This AM superposition splits into several solitary waves, which are localised around individual fibre cores (panels \textbf{e)} and \textbf{f)}). Two such solitary waves which are completely formed are highlighted by blue arrows; the third trailing pulse in this train will split into another pair of solitary waves localised on $n=1$ and $n=2$ fibre cores respectively.}}  
\label{fig: symmbreak}
\end{figure}

In this section a small but non-negligible third order dispersion coefficient is used, $\beta_3 = 1/10$. To facilitate rapid spectral broadening, we choose as the input condition to \eqref{eq: coupledNLSE} an intense pulse ($A_n(0, t) = 4 \overline{A}_n(0,t)$, $B=1$ $\forall$ $n=1,2,...N$) which has energy well in excess of that required for the excitation of a single soliton. Within a short propagation distance, we observe the input pulse split into several components with different group velocities. These fission dynamics are typical of those experienced by higher-order solitons in single fibres \cite{Husakou2001} and do not necessarily induce symmetry breaking. In the absence of noise, AM different from that of the pump $l \neq m$ may remain unpopulated for a considerable distance after the initial fission process is complete. However, the brightest emerging pulse will typically experience growth of other $l \neq m$ AM around its peak, which destabilise it and cause a second fission event. The propagation distance required to observe this second fission is usually much shorter for non-zero input winding numbers $m$ in the absence of fibre twisting. The fact that angular symmetry is always broken at the point in time with the highest intensity demonstrates that nonlinear cross-phase modulation terms between different AM are primarily responsible for the instability. After the second fission a train of pulses form with fluctuating populations across the AM spectrum, releasing dispersive waves as these beat against each other. The oscillations slowly relax leaving each pulse with a roughly equal distribution among all the available AM, which corresponds to localisation around single fibre cores. These spatio-temporal solitary waves propagate stably thereafter and are similar to the discrete spatio-temporal solitons previously observed in one-dimensional straight waveguide arrays \cite{Babushkin2007}. This process is illustrated for a six-core ($N=6$) fibre with modest coupling $\Delta = 1$ pumped with an $m=0$ input pulse in figure \ref{fig: symmbreak}. 

\begin{figure}[h] 
\centering
\includegraphics[width=\linewidth]{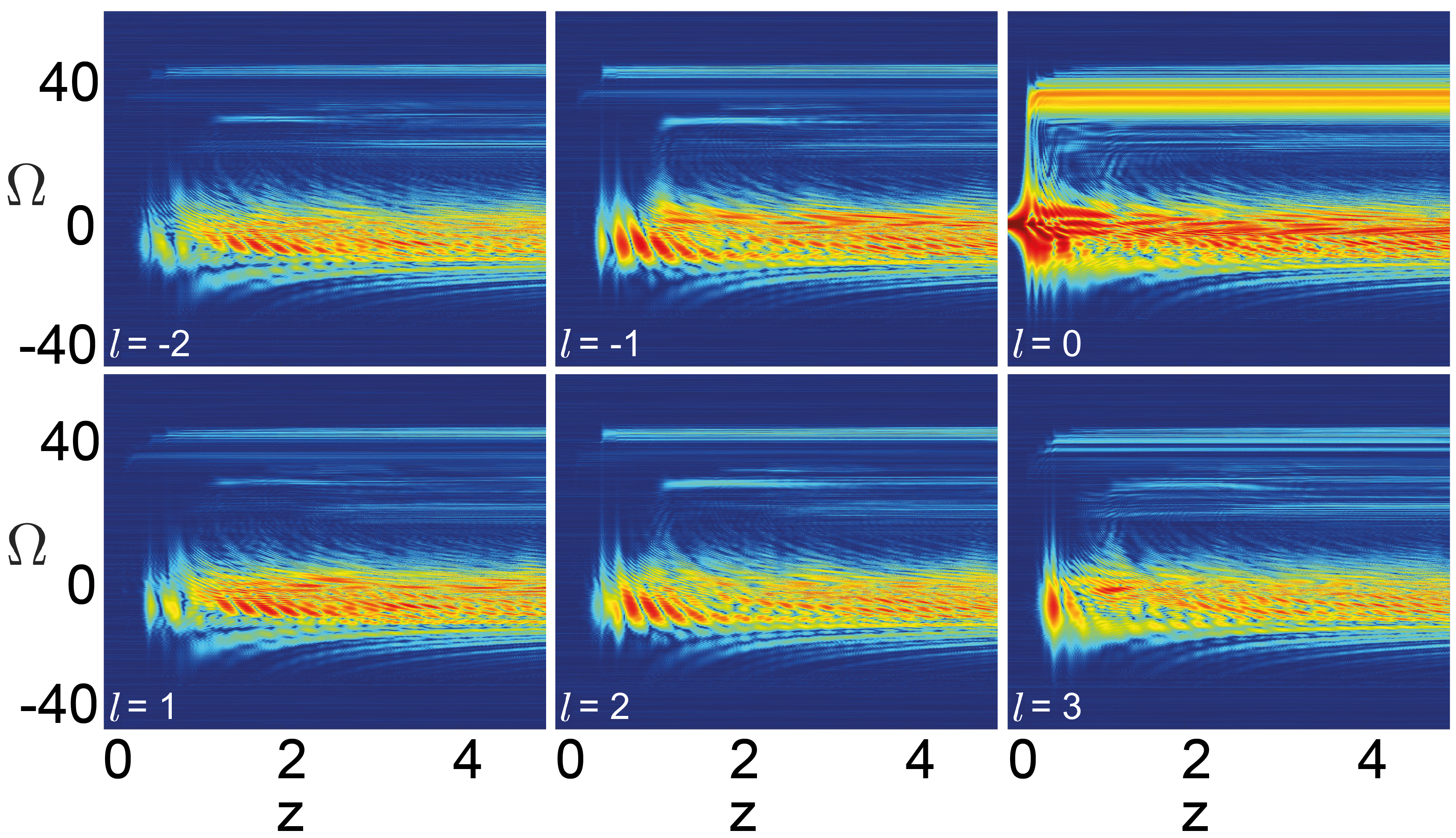}
\caption{\small{Frequency $\Omega$ spectra vs. propagation distance $z$ for different AM $l$, resulting from the pulse evolution over five dispersion lengths shown in figure \ref{fig: symmbreak}.}}  
\label{fig: m0spectra}
\end{figure}

\begin{figure}[h] 
\centering
\includegraphics[width=\linewidth]{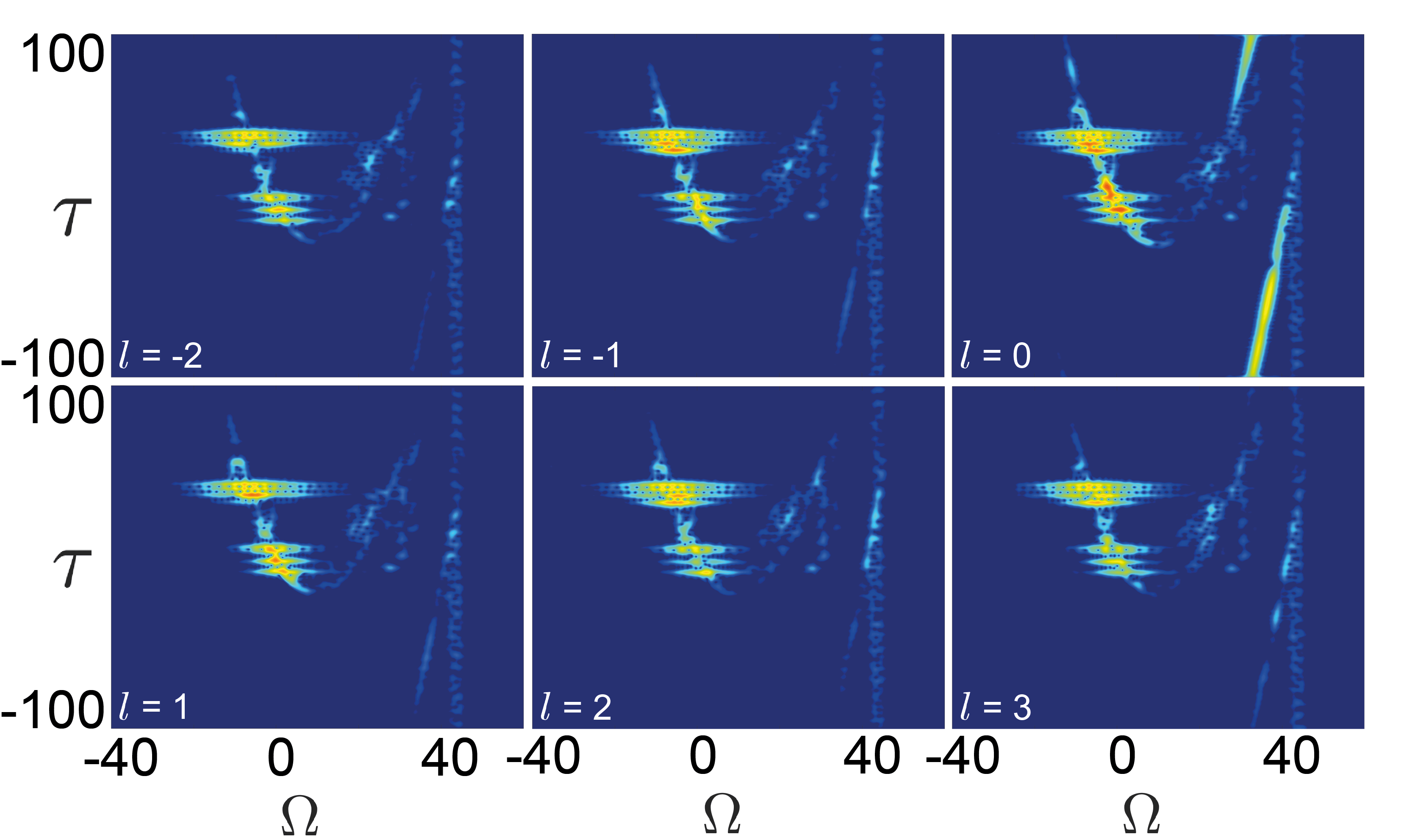}
\caption{\small{XFROG maps $S_l(\tau, \Omega)$ resolved for each AM $l$ resulting from the pulse propagation shown in figure \ref{fig: symmbreak}, evaluated at $z=5L_D$. }}  
\label{fig: m0xfrog}
\end{figure}

The frequency spectrum may be resolved for each AM by a continuous Fourier transform in time, followed by a discrete transform across the fibre index $n$,
\begin{equation} \label{eq: spectrumdecomp}
\begin{split}
A_l(z, \Omega) = &\frac{1}{\sqrt{N}} \sum_{n=1}^{N} \exp\left(-i \frac{2 \pi l n}{N}\right)\\ 
&\times \int_{-\infty}^{\infty} dt \exp\left(- i \Omega t\right) A_n(z, t).
\end{split}
\end{equation} 
The dimensionless frequency $\Omega$ expresses the relative detuning from the pump's central frequency $\Omega=0$. The spectra resulting from the propagation shown in figure \ref{fig: symmbreak} are plotted on a logarithmic scale in figure \ref{fig: m0spectra}, with each sub-figure showing the spectral evolution for a different AM $l$, with the pump AM being $m=0$. The growth of non-pump AM $l \neq m = 0$ is initiated by angular symmetry breaking just after the spectral broadening in $l=0$ reaches its maximum extent, seeding a band of dispersive radiation around $\Omega = 35$. Here the dispersive radiation is caused by the presence of $\beta_{3}$ in eq. (\ref{eq: coupledNLSE}).  These non-pump AM in turn generate their own dispersive bands, whose strength and central frequency is generally $l$-dependent. 
To get a better insight into the dynamics of supercontinuum generation, we perform an XFROG trace which lets us observe the pulse across time and frequency domains simultaneously \cite{Efimov2005}. Taking the input pump pulse profile as the temporal correlation function, we define the XFROG map as \cite{Agrawal2013} 
\begin{equation} \label{eq: xfrogdef}
S_l(\tau, \Omega) = {\left|\int_{-\infty}^{\infty} A_l(z, t) \sech(t-\tau) \exp(-i \Omega t) dt  \right|}^2
\end{equation} 
noting with the $l$ subscript in $A_l(z, t)$ that we have already performed the discrete Fourier transform over $n$ into the AM basis. Figure \ref{fig: m0xfrog} shows the XFROG maps associated with the fields shown in figure \ref{fig: symmbreak} \textbf{f)}. It is clear from this that the bulk of the dispersive radiation carries no AM (being mainly present in the $l=0$ panels) as it is generated during the initial stages of fission, before  angular symmetry is broken.

\section{Twisting Effects}

\begin{figure}[h] 
\centering
\includegraphics[width=\linewidth]{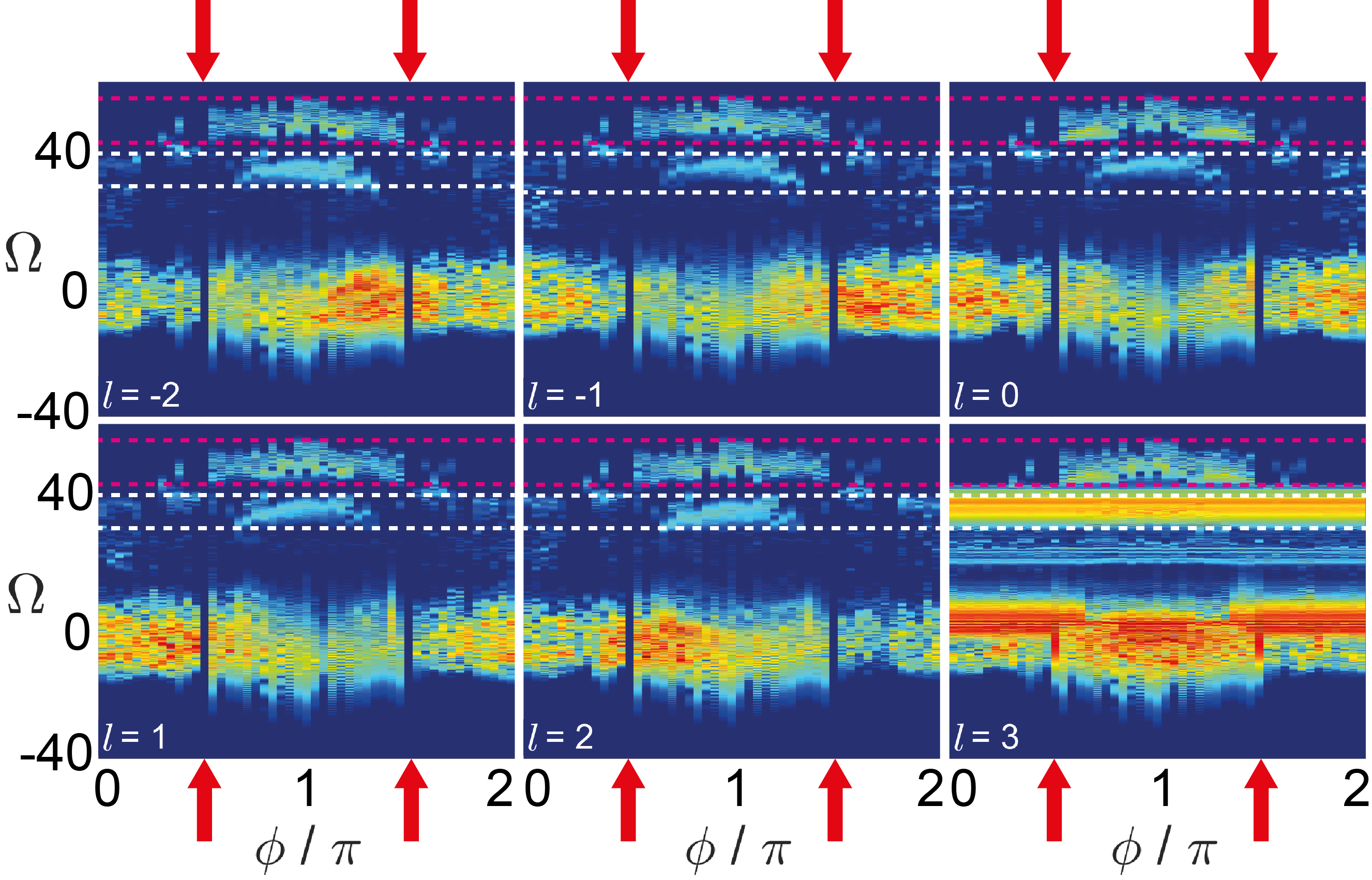}
\caption{\small{Output spectra resolved by AM $l$ after propagating an intense ($A_n(0, t) = 4 \overline{A}_n(0,t)$, $B=1$) pump pulse carrying AM $m=3$ for $z=5L_D$ in a strongly coupled array $\Delta=10$, given varying values of the twist-induced Peierls phase $\phi$. No light is generated in $l \neq m$ channels when $\phi=\pi/2, 3\pi/2$ as indicated by the red arrows, due to the symmetry breaking suppression mechanism described above. Two bands of dispersive radiation are indicated by dashed white and magenta lines in each $l$ subfigure, which are excited during the splitting of the initial pulse and a post-fission pulse respectively.}}  
\label{fig: phispectra}
\end{figure}

\begin{figure}
\centering
\includegraphics[width=\linewidth]{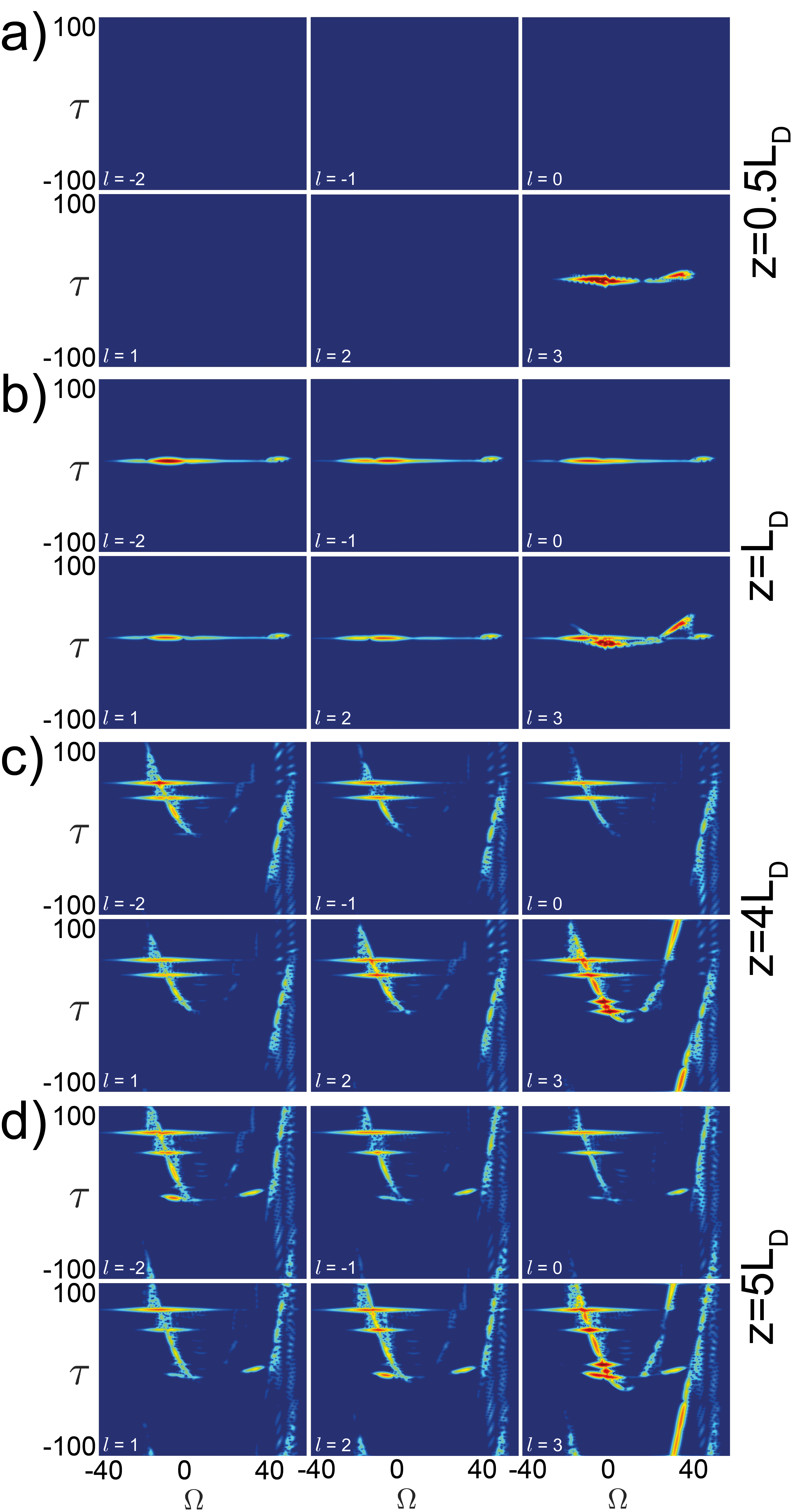}
\caption{\small{XFROG maps $S_l(\tau, \Omega)$ of the evolution of an an intense ($A_n(0, t) = 4 \overline{A}_n(0,t)$, $B=1$) input pulse with $m=3$ in a strongly coupled $\Delta=10$ ring with $\phi=\pi$, evaluated after propagation distances of \textbf{a)} $z=0.5L_D$,  \textbf{b)} $z=L_D$,  \textbf{c)} $z=4L_D$,  \textbf{d)} $z=5L_D$. The initial soliton fission in \textbf{a)} occurs prior to any symmetry breaking and does not populate any of the $l \neq m$ AM channels. Symmetry breaking occurs when the leading post-fission pulse first breathes and broadens in frequency \textbf{b)}, exciting a resonance around $\Omega=45$. Both this and the post-fission pulse train couple to dispersive waves, acquiring a time delay as they propagate \textbf{c)}. Eventually the trailing $l=3$ pulse experiences symmetry breaking \textbf{d)} and induces another resonance around $\Omega=35$. }}  
\label{fig: m3PhiPiXFrogs}
\end{figure}

The Peierls phase $\phi$ introduced by the fibre ring twist has a great impact on the output supercontinuum. In particular it appears that certain values of $\phi$ can stabilise input pulses with a single AM, allowing for spectral broadening without breaking of angular symmetry and generating other AM. The stabilising value of $\phi$ depends both on the initial AM $m$ and the number of fibre cores $N$.  Viewing the symmetry breaking at the pulse's peak intensity in time as a kind of `instantaneous modulation instability', we find that to first order symmetry breaking should be prevented when $\cos(2\pi m/N - \phi) = 0$, which occurs for two distinct $\phi$ in the interval $(0, 2\pi)$. We illustrate this in figure \ref{fig: phispectra} given an $m=3$ pump pulse and $N=6$; in this case the spectra for $\phi = \pi/2, 3\pi /2$ (highlighted by red arrows) show precisely zero occupation in AM $l \neq m =3$ as for these Peierls phases angular symmetry is preserved.
Besides the complex substructure of peaks within the spectrally-broadened envelope around the pump frequency $\Omega=0$, two dispersive bands can be seen in the spectrum of each AM. These bands appear predominantly for $|\phi-\pi|<\pi/2$ and result from two symmetry breaking events. To demonstrate this we examine XFROG traces at different stages of propagation for the $\phi=\pi$ case, shown in figure \ref{fig: m3PhiPiXFrogs}. The first band (lying between dashed white lines in each subfigure) is created due to the leading pulse splitting post-fission (Fig.~ \ref{fig: m3PhiPiXFrogs}\textbf{b)}, while the second (between dashed magenta lines) is due to the trailing post-fission pulse destabilising \textbf{d)}. Since the trailing pulse is much less intense than the leading one, azimuthal instability takes longer to set in and the resonant frequency is red-shifted in comparison to the first band. Conversely, in the $l=m=3$ spectrum, there is a large component which is independent of $\phi$ since this is generated by the initial soliton fission prior to angular symmetry breaking. 
\clearpage

\begin{figure}
\centering
\includegraphics[width=0.6\linewidth]{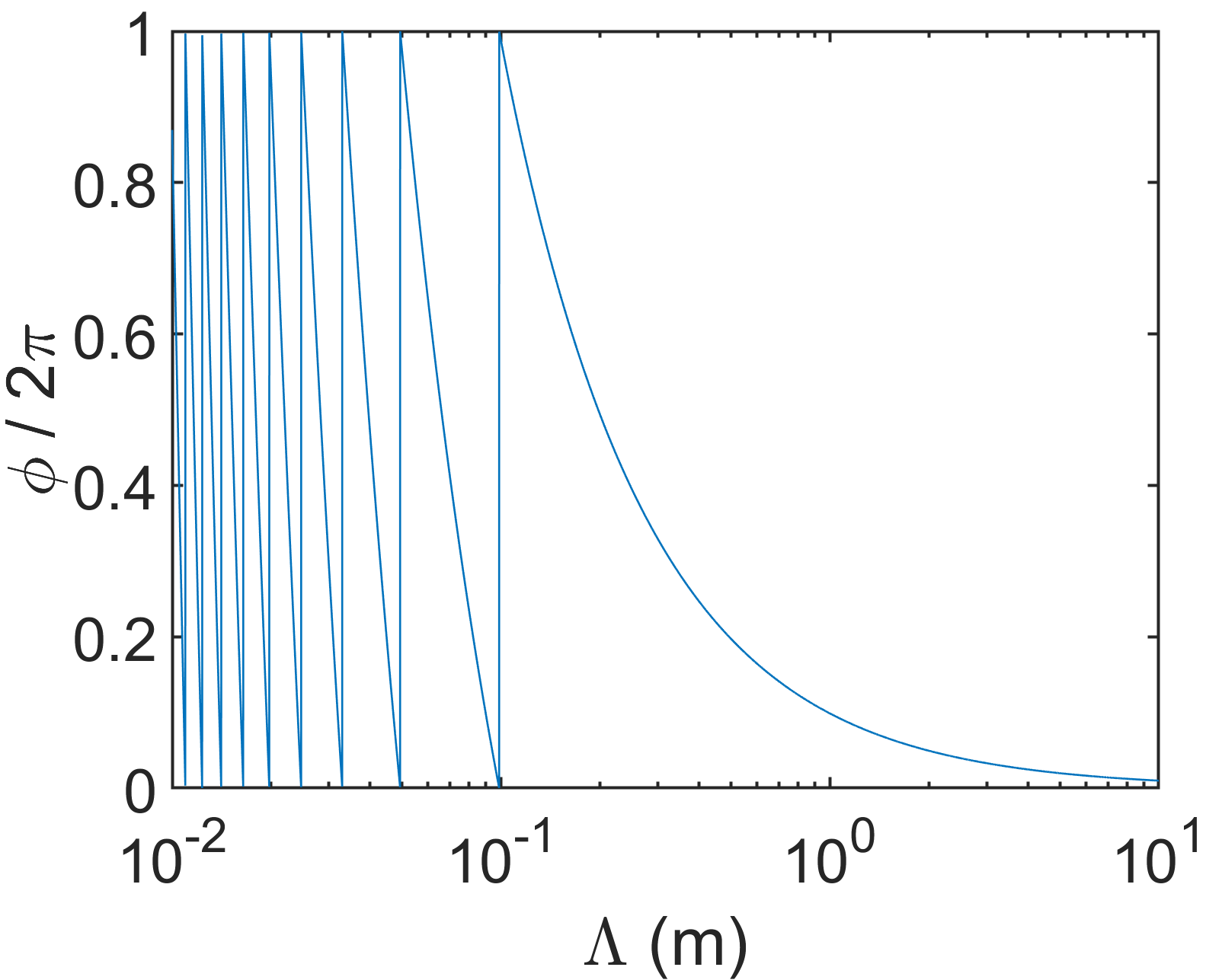}
\caption{\small{Plot of the Peierls phase $\phi$ as a function of twist period $\Lambda$ as per eq. \eqref{eq: twistphase}, assuming $n_s=1.5$, $r_0=100\mu$m and $\lambda_0=1\mu$m.}}  
\label{fig: phivslambda}
\end{figure}

We would like to point out that the full range of physically significant $\phi$ should be experimentally accessible for a suitably designed fibre core array. $\phi$ is related to the fibre ring twist period $\Lambda$ through \cite{Longhi2007b} 
\begin{equation} \label{eq: twistphase}
\phi =  \frac{8 \pi^3 n_s {r_0}^2}{N \lambda_0 \Lambda}
\end{equation}
where $n_s$ is the substrate refractive index and $r_0$ is approximately the ring radius. Plotting this for realistic parameters (figure \ref{fig: phivslambda}) suggests all possible Peierls phases from $0$ to $2\pi$ can be realised with twist periods $\Lambda \geqslant 10$cm, which should be sufficiently mild as to avoid excessive fibre bending losses.

\section{Coupling Dependent Resonant Radiation}

\begin{figure}[h] 
\centering
\includegraphics[width=\linewidth]{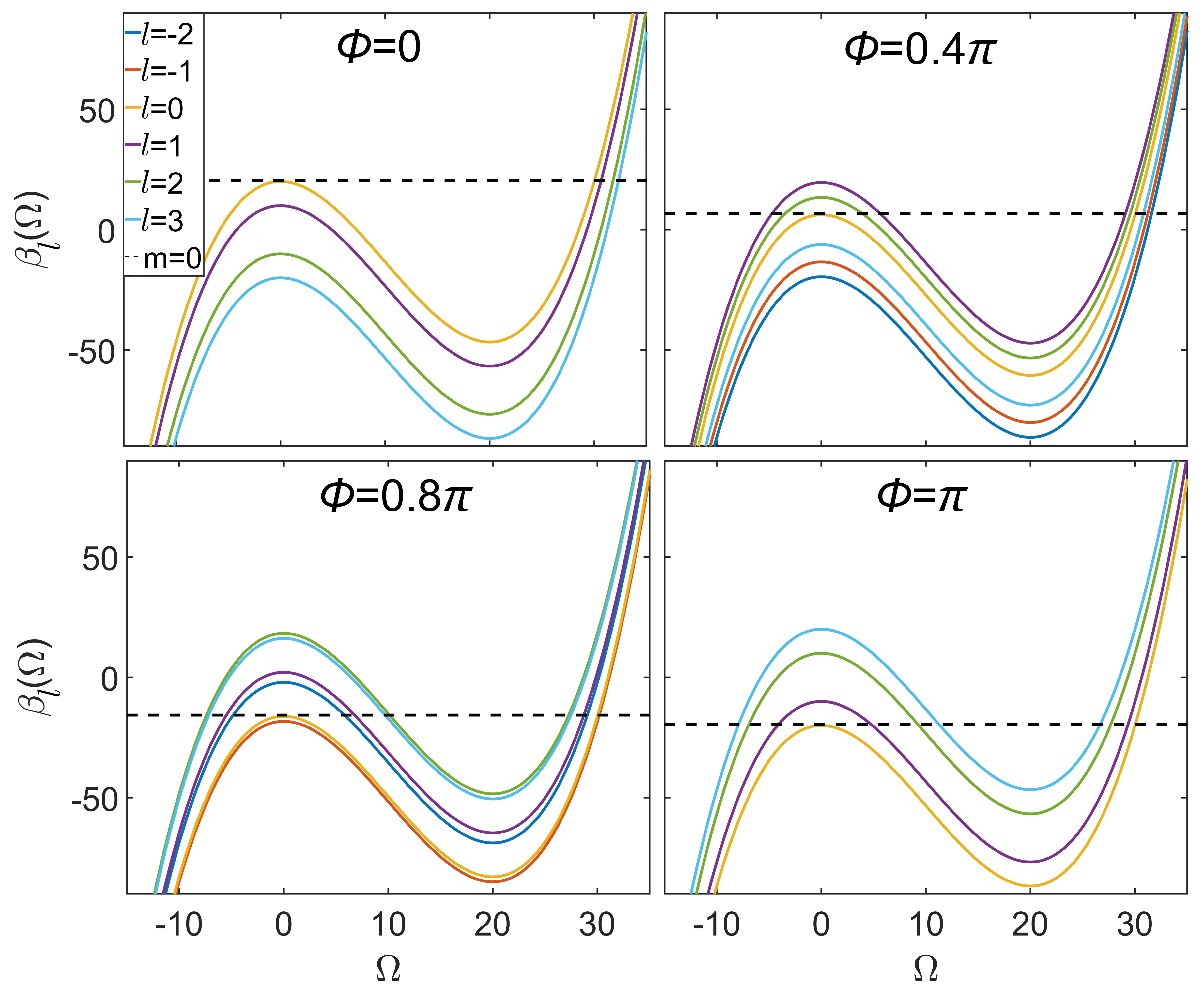}
\caption{\small{Propagation constants for dispersive waves $\beta_l(\Omega)$ for comparison with that of a soliton without AM ($m=0$, black dashed line), assuming various values of the Peierls phase $\phi$. Here we have taken $B=1$, $\beta_2 <0$, $\beta_3 = -1/10$ and $\Delta=10$. }}  
\label{fig: phasematching}
\end{figure}

\begin{figure}[h] 
\centering
\includegraphics[width=\linewidth]{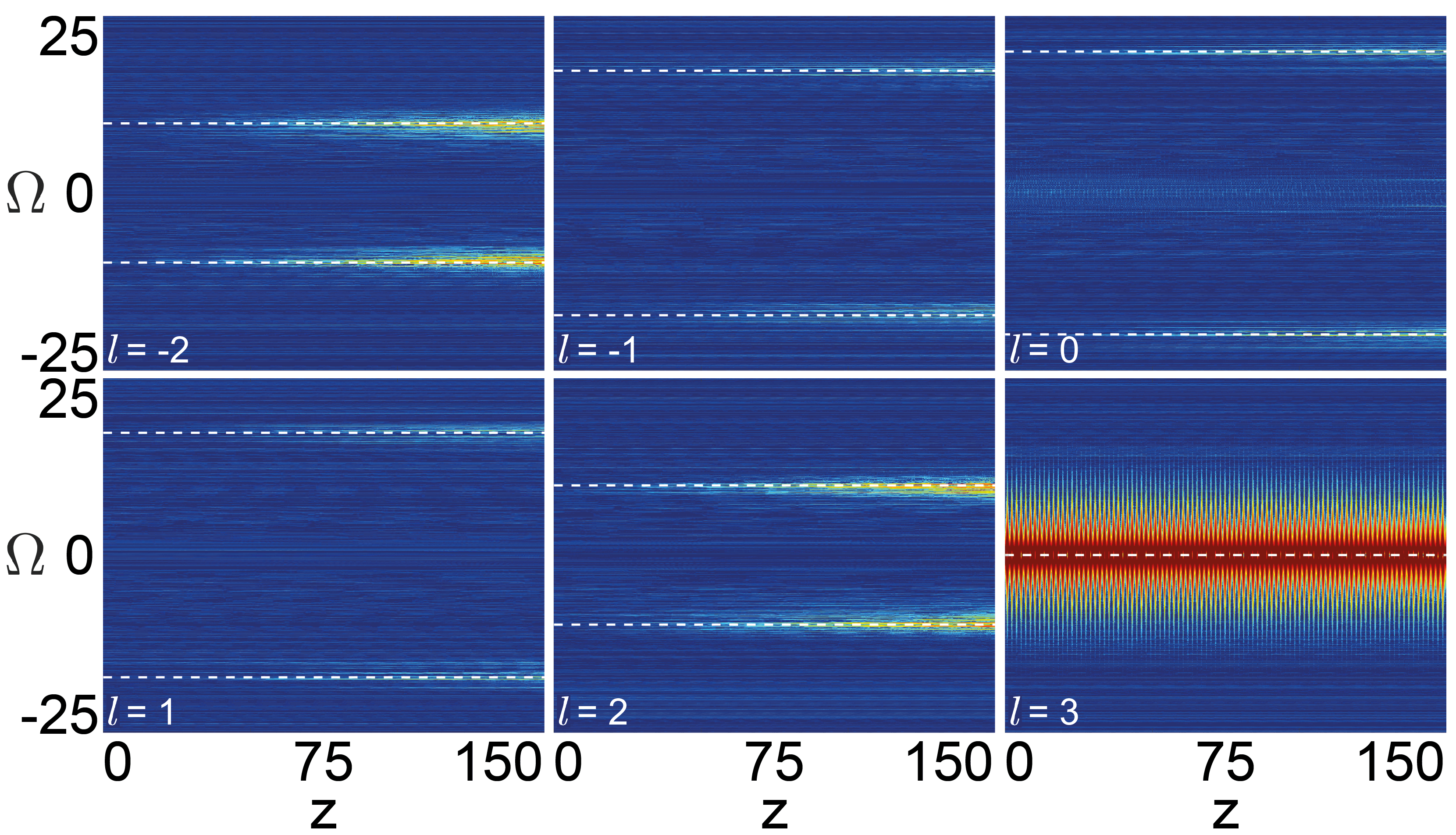}
\caption{\small{AM resolved spectra resulting from propagation of an input pulse $A_n(0, t) = 2 \overline{A}_n(0,t)$, $B=1$ with AM $m=3$ over $150$ dispersion lengths, through a strongly coupled $\Delta=50$ untwisted $\phi=0$ six-core array, with only second-order anomalous dispersion $\beta_2 <0$, $\beta_3=0$. White dashed lines indicate the resonant frequencies $\Omega_{RR}(l)$ predicted by equation \eqref{eq: quadomegaRR}. A high coupling strength $\Delta=50$ is used here.}}  
\label{fig: resrad}
\end{figure}

Resonant radiation occurs when the pump pulse becomes phase matched with dispersive waves in the fibre's continuum, meaning they have the same propagation constant. Comparing the wavenumber $\beta_l$ of plane wave solutions $\propto \exp(i \left( \beta_l z + l \theta_n - \Omega t \right) )$ to equation  \eqref{eq: coupledNLSE} (neglecting the nonlinearity) with that of the soliton solution $\overline{\beta} _m$ provides a phase matching condition. Assuming anomalous dispersion $\beta_2<0$, the phase matching condition is
\begin{equation} \label{eq: phasematch}
\begin{split}
-\frac{1}{2} \Omega^2 +  \frac{\beta_3}{6} \Omega^3 + &2 \Delta \cos( \frac{2 \pi l}{N} - \phi)\\
= \frac{B^2}{2} + &2 \Delta \cos( \frac{2 \pi m}{N}  - \phi). 
\end{split}
\end{equation}
This cubic equation for the resonant frequency $\Omega$ has analytic, albeit complicated, solutions. Plotting the dispersion of both the dispersive waves and the discrete vortex soliton (Fig.~\ref{fig: phasematching}) shows that up to three of these solutions may be real for $l \neq m$, meaning multiple resonant radiation frequencies may be excited for a single AM. So long as $\beta_3>0$ there is always at least one resonant frequency per AM (the largest possible), which is the same as would appear in a single fibre core. Whether the two other resonances may be excited depends on the combination of Peierls phase $\phi$ applied, pump ($m$) and resonance AM ($l$),  as well as the coupling strength $\Delta$. 
\begin{figure}[h] 
\centering
\includegraphics[width=0.6\linewidth]{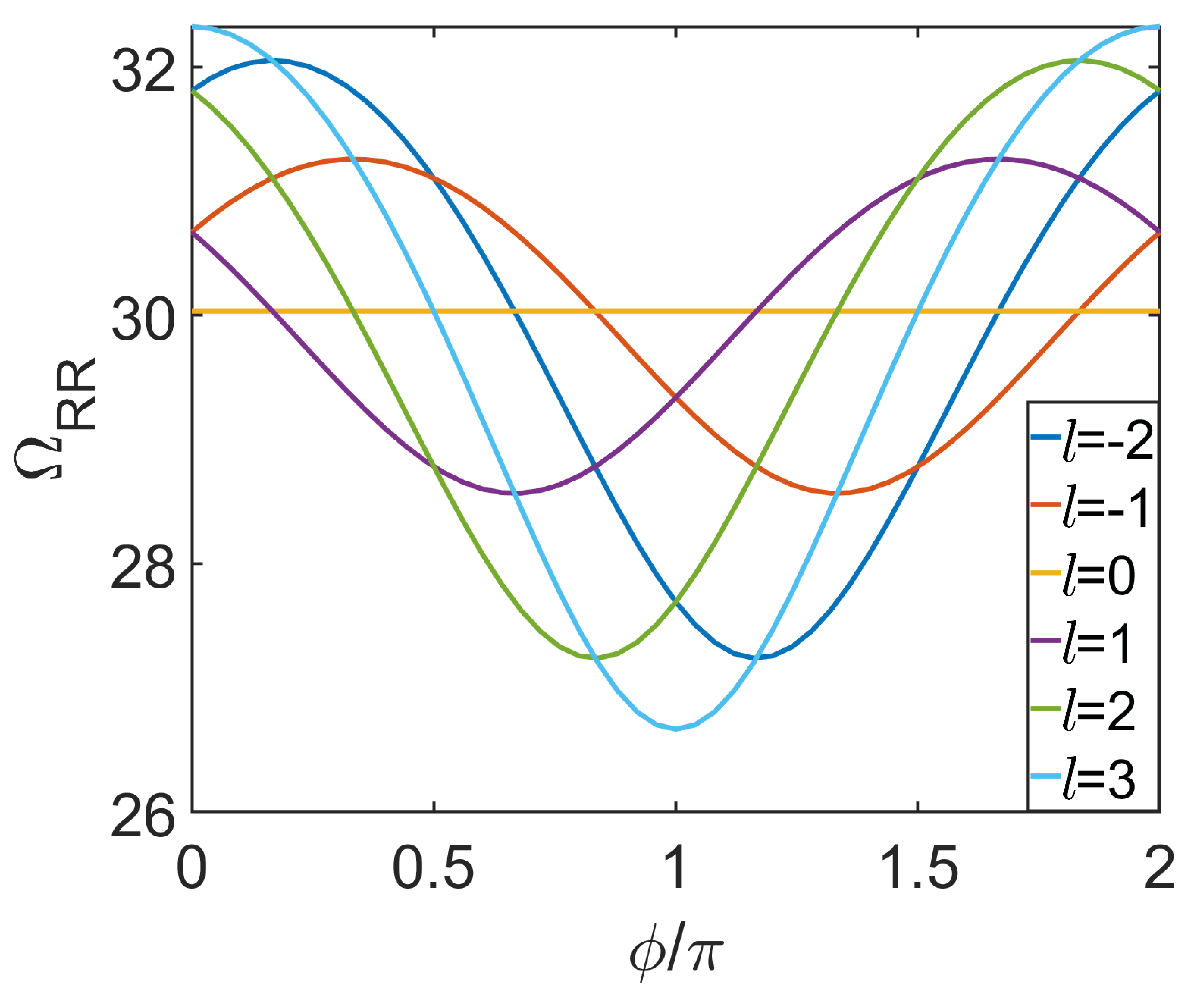}
\caption{\small{Frequency of `standard' resonant radiation as can appear in single fibres, which is always present while $\beta_3>0$, as a function of Peierls phase $\phi$ given the parameters in Fig.~\ref{fig: phasematching}. Note that for the pump's AM $l=m=3$ this is independent of $\phi$ and is identical to the resonant frequency which would be observed in a single uncoupled fibre.}}  
\label{fig: omegaRRplot}
\end{figure}
In the absence of third order dispersion Eq. \ref{eq: phasematch} is quadratic with a pair of solutions symmetric about the pump frequency $\Omega=0$:
\begin{equation} \label{eq: quadomegaRR}
\Omega_{RR} = \pm \sqrt{4 \Delta \left(\cos( \frac{2 \pi l}{N}  - \phi) - \cos( \frac{2 \pi m}{N}  - \phi)\right) - B^2  } 
\end{equation}
In single fibres resonant radiation cannot appear without higher order dispersion. However in this case phase matching for non-pump AM is possible providing the coupling rate is comparable to the pump intensity, meaning for the dimensionless parameters used here $\Delta \geqslant 1 / 4$. This kind of coupling-enabled radiation has been reported in other coupled fibre systems \cite{Benton2008, Oreshnikov2017} and is always present in multicore structures due to spatial discreteness. In figure \ref{fig: resrad} we show the simulated spectral evolution of a bright pulse $A_n(0, t) = 2 \overline{A}_n(0,t)$, $B=1$ in an untwisted six-core fibre ring with $\beta_2 <0$, $\beta_3=0$; sidebands develop around the predicted frequencies for each AM $l \neq m$.

\section{Conclusion}

We have shown how supercontinua with a wide distribution of AM as well as frequencies may be realised in a twisted fibre array. Instabilities driven by four-wave mixing between different AM cause spontaneous breaking of angular symmetry, resulting in a single-AM input pulse splitting into a train of solitary waves localised both in time and the discrete angular coordinate. Controlling the effective Peierls phase with the array twist period allows the supercontinuum spectrum to be manipulated, possibly restricting it to a single AM by stabilising the pump against nonlinear instabilities. Coupling between fibre cores modifies the dispersion of different AM, enabling new resonant radiation modes through additional phase matching conditions which are not possible in uncoupled fibres. They may occur without intrinsic higher-order dispersion coefficients, as they arise due to the lattice dispersion of the periodic ring array of cores. Twisting the fibre ring also allows these resonant frequrncies to be tailored. Our results could be also be useful to researchers working on twisted solid-core and hollow-core photonic crystal fibres \cite{Russel2017, Roth2018} and the development of supercontinua in such structures. 

C.M. acknowledges studentship funding from EPSRC under CM-CDT Grant No. EP/L015110/1. F.B. acknowledges support from the German Max Planck Society for the Advancement of Science (MPG), in particular the IMPP partnership between Scottish Universities and MPG.

\bibliography{TwistedFiberIdeasBib}

\end{document}